\documentclass[epjST]{svjour}
\usepackage[T1]{fontenc}
\usepackage[latin9]{inputenc}
\usepackage{mathrsfs}
\usepackage{amsmath}
\usepackage{amssymb}
\usepackage{float}
\usepackage{graphicx}
\usepackage{esint}
\usepackage{array}

\begin{document}

\title{Synchronization of two self-excited double pendula }

\author{Piotr Koluda\inst{1} \and Przemyslaw Perlikowski\inst{1} \fnmsep\thanks{\email{przemyslaw.perlikowski@p.lodz.pl}} \and Krzysztof Czolczynski\inst{1} \and Tomasz Kapitaniak\inst{1}}
\institute{Division of Dynamics, Faculty of Mechanical Engineering, Lodz University of Technology, 90-924 Lodz, Stefanowskiego 1/15, Poland}

\abstract{
We consider the synchronization of two self-excited double pendula.
We show that such pendula hanging on the same beam can have four
different synchronous configurations. Our approximate analytical analysis
allows us to derive the synchronization conditions and explain the observed
types of synchronization. We consider an energy balance in the system
and describe how the energy is transferred between the pendula via the
oscillating beam, allowing thus the pendula  synchronization. Changes
and stability ranges of the obtained solutions with increasing and decreasing
masses of the pendula are shown using path-following. 
}
\maketitle

\section{Introduction}

Synchronization is commonly observed to occur among oscillators
\cite{ref1,ref3,ref15,ref21,ref24}. It is a process where two
or more systems interact with one another and come to oscillate together.
Groups of oscillators are observed to synchronize in a diverse variety
of systems, despite inevitable differences between oscillators.
The history of synchronization goes back to the 17th century. In 1673
the Dutch scientist Ch. Huygens observed weak synchronization of two pendulum
clocks \cite{ref9}. Recently, the phenomenon of synchronization
of clocks hanging on a common movable beam \cite{ref10} has
been the subject of research conducted by numerous authors \cite{ref9,ref2,ref4,ref5,ref6,ref7,ref8,ref11,ref13,ref14,ref17,ref19}.
These studies explain the phenomenon of synchronization of a number
of single pendula.

In our work we consider an interaction between two double pendula.
One of the first investigations on dynamics of the double pendulum
can be found in the paper by Rott \cite{ref35}, where an analytical
investigation of the Hamiltonian system for different ratios between natural
frequencies of pendula is presented. The next results obtained by
Miles \cite{ref29} show dynamics of the double pendulum under parametric
excitation around the $2:1$ resonance. A mode interaction in the double
pendulum, including a detailed bifurcation analysis near two multiple
bifurcation points and a transition to quasi-periodic motion and chaos
around the $2:1$ parametric resonance, is presented in \cite{ref28,ref27,ref26}.
Similarly as for $2:1$, the $1:1$ resonance leads to dynamics
including multiple bifurcation points, symmetry breaking and cascades
of period doubling bifurcations \cite{ref26}. Double pendula can
be also considered as an example of many physical systems commonly met
in engineering, e.g., a model of bridge-pedestrian interactions \cite{ref31},
golf or hockey swing interactions with arms \cite{ref34}, human
body \cite{ref33} or trunk \cite{ref32} models.

In this paper we consider the synchronization of two self-excited
double pendula. The oscillations of each double pendulum are self-excited
by the van der Pol type of damping associated with the upper parts (upper
pendula) of each double pendulum. We show that two such double pendula
hanging on the same beam can synchronize both in phase and in anti-phase.
We give an evidence that the observed synchronous states are robust
as they exist for a wide range of system parameters and are preserved
for the parameter mismatch. The performed approximate analytical
analysis allows us to derive the synchronization conditions and explain
the observed types of synchronization. The energy balance in the
system allows us to show how the energy is transferred between the pendula
via the oscillating beam.

This paper is organized as follows: Section 2 describes the considered
model of the coupled double pendula, in Section 3 we derive an energy
balance of the synchronized pendula, whereas Section 4 presents the results of
our numerical simulations and describes the observed synchronization
states and their ranges of stability. Finally, we summarize our results
in Section 5.

\section{Model \label{sec:Model}}

The analyzed system is shown in Fig. 1. It consists of a rigid beam
and two double pendula suspended on it. The beam of the mass $M$ can
move along the horizontal direction, its movement is described by the coordinate
$x_{b}$.The beam is connected to a linear spring and a linear damper, $k_{x}$ and $c_{x}$.

\begin{figure}[!ht]
\begin{centering}
\includegraphics{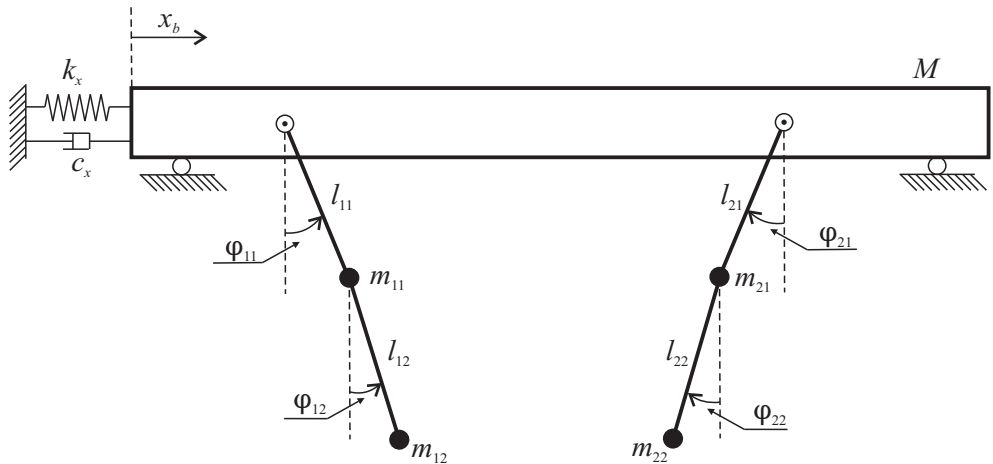}
\caption{Model of the system - two double pendula are mounted to the beam
which can move horizontally. Each double pendulum consists of an
upper pendulum of the length $l_{i1}$ and the mass $m_{i1}$ and a lower
pendulum of the length $l_{i2}$ and the mass $m_{i2}$ ($i=1,2$). The
upper pendula are self-excited. }

\par\end{centering}

\label{fig:1} 
\end{figure}

Each double pendulum consists of two light beams of the length $l_{i1}$
and the masses $m_{i1}$ ($i$-th upper pendulum) and the length $l_{i2}$
and $m_{i2}$ ($i$-th lower pendulum), where $i=1,2$, mounted
at its ends. We consider double pendula with the same lengths
$l_{11}=l_{21}=l_{12}=l_{22}=l$ but different masses $m_{i1}$ and
$m_{i2}$ (to maintain generality in the derivation of equations, we use indexes
for lengths of the pendula). The motion of each double pendulum is described
by the angles $\varphi_{i1}$ (upper pendulum) and $\varphi_{i2}$ (lower
pendulum). The upper pendula are self-excited by the van der Pol type
of damping (not shown in Fig. 1) given by the momentum (torque) $c_{vdp}\dot{\varphi}_{1i}(1-\mu\varphi_{1i}^{2})$,
where $c_{vdp}$ and $\mu$ are constant. Van der Pol damping results
in the generation of a stable limit cycle \cite{ref1}. The lower pendula
are damped with a viscous damper with the coefficient $c_{i2}$. The equations
of motion of the considered system are as follows: 
\begin{equation}
\begin{array}{lcr}
(M+\sum_{i=1}^{2}\sum_{j=1}^{2}m_{ij})\ddot{x}_{b}+\sum_{i=1}^{2}(m_{i1}+m_{i2})l_{i1}(\ddot{\varphi}_{i1}\cos\varphi_{i1}-\dot{\varphi}_{i1}^{2}\sin\varphi_{i1})+\\
\sum_{i=1}^{2}m_{i2}l_{i2}(\ddot{\varphi}_{i2}\cos\varphi_{i2}-\dot{\varphi}_{i2}^{2}\sin\varphi_{i2})+k_{x}x_{b}+c_{x}\dot{x}_{b}=0\\
(m_{i1}+m_{i2})l_{i1}\ddot{x}_{b}\cos\varphi_{i1}+(m_{i1}+m_{i2})l_{i1}^{2}\ddot{\varphi}_{i1}+m_{i2}l_{i1}l_{i2}\ddot{\varphi}_{i2}\cos(\varphi_{i1}-\varphi_{i2})+\\
m_{i2}l_{i1}l_{i2}\dot{\varphi}_{i2}^{2}\sin(\varphi_{i1}-\varphi_{i2})+(m_{i1}+m_{i2})l_{i1}g\sin(\varphi_{i1})+c_{vdp}(1-\mu\varphi_{i1}^{2})\dot{\varphi}_{i1}+c_{i2}(\dot{\varphi}_{i2}-\dot{\varphi}_{i1})=0\\
m_{i2}l_{i2}\ddot{x}_{b}\cos\varphi_{i2}+m_{i2}l_{i1}l_{i2}\ddot{\varphi}_{i1}\cos(\varphi_{i1}-\varphi_{i2})+m_{i2}l_{i2}^{2}\ddot{\varphi}_{i2}\\
-m_{i2}l_{i1}l_{i2}\dot{\varphi}_{i1}^{2}\sin(\varphi_{i1}-\varphi_{i2})+m_{i2}l_{i2}g\sin(\varphi_{i2})-c_{i2}(\dot{\varphi}_{i2}-\dot{\varphi}_{i1})=0.
\end{array}\label{eq:DimForm}
\end{equation}
Introducing the dimensionless time $\tau=\omega t$, where $\omega^{2}=\frac{g}{l_{11}}$
is the natural frequency of the upper pendula, we can rewrite Eq.
(\ref{eq:DimForm}) in the dimensionless form as: 
\begin{gather}
\ddot{y}_{b}+\sum_{i=1}^{2}\mathbf{A}_{i1}(\ddot{\psi}_{i1}\cos\psi_{i1}-\dot{\psi}_{i1}^{2}\sin\psi_{i1})+\sum_{i=1}^{2}\mathbf{A}_{i2}(\ddot{\psi}_{i2}\cos\psi_{i2}-\dot{\psi}_{i2}^{2}\sin\psi_{i2})+\label{eq:2:a}\\
+\mathbf{K}y_{b}+\mathbf{C}\dot{y}_{b}=0\nonumber \\
\delta_{i1}\ddot{y}_{b}\cos\psi_{i1}+\mathbf{L}_{i1}\ddot{\psi}_{i1}+\textbf{L}_{i3}\ddot{\psi}_{i2}\cos(\psi_{i1}-\psi_{i2})=\label{eq:2:b}\\
-\textbf{L}_{i3}\dot{\psi}_{i2}^{2}\sin(\psi_{i1}-\psi_{i2})-\textbf{G}_{i1}\sin(\psi_{i1})-\mathbf{C}_{vdp}(1-\mu\psi_{i1}^{2})\dot{\psi}_{i1}-\mathbf{C}_{i2}(\dot{\psi}_{i2}-\dot{\psi}_{i1})\nonumber \\
\delta_{i2}\ddot{y}_{b}\cos\psi_{i2}+\textbf{L}_{i3}\ddot{\psi}_{i1}\cos(\psi_{i1}-\psi_{i2})+\textbf{L}_{i2}\ddot{\psi}_{i2}=\label{eq:2:c}\\
\textbf{L}_{i3}\dot{\psi}_{i1}^{2}\sin(\psi_{i1}-\psi_{i2})-\textbf{G}_{i2}\sin(\psi_{i2})+\mathbf{C}_{i2}(\dot{\psi}_{i2}-\dot{\psi}_{i1})\nonumber 
\end{gather}
where $\mathbf{A}_{i1}=\frac{(m_{i1}+m_{i2})l_{i1}}{{M}l_{b}}$, $\mathbf{A}_{i2}=\frac{m_{i2}l_{i2}}{{M}l_{b}}$,
$\mathbf{K}=\frac{k_{x}}{{M}\omega^{2}}$, $\textbf{C}=\frac{c_{x}}{{M}\omega}$,
$\delta_{i1}=\frac{(m_{i1}+m_{i2})l_{i1}}{{M}l_{12}}$, $\delta_{i2}=\frac{m_{i2}l_{i2}}{{M}l_{12}}$,
$\textbf{L}_{i1}=\frac{(m_{i1}+m_{i2})l_{i1}^{2}}{l_{12}l_{b}{M}}$,
$\textbf{L}_{i2}=\frac{m_{i2}l_{i2}^{2}}{l_{12}l_{b}{M}}$, $\textbf{L}_{i3}=\frac{m_{i2}l_{i2}l_{i1}}{l_{b}{M}l_{12}}$,
$\textbf{G}_{i1}=\frac{(m_{i1}+m_{i2})l_{i1}g}{l_{12}\omega^{2}l_{b}{M}}$,
$\textbf{G}_{i2}=\frac{m_{i2}l_{i2}g}{l_{12}\omega^{2}l_{b}{M}}$,
$\mathbf{C}_{vdp}=\frac{c_{vdp}}{\omega l_{b}{M}l_{12}}$, $\mathbf{C}_{i2}=\frac{c_{i2}}{l_{12}\omega l_{b}{M}}$.

\section{Analytical conditions for synchronization \label{sec:Analitical-solution}}

\subsection{Force with which the pendula act on the beam}

In this section we derive an approximate analytical condition for
the pendulum synchronization in the considered system. Assuming that
the double pendula are identical and perform periodic oscillations
with the frequency $\omega_{0}$ and low amplitudes, one can describe
displacements, velocities and accelerations of the upper and lower
pendula in the following way: 
\begin{align}
\psi_{ij}=\Phi_{ij}\sin(\omega_{0}\tau+\beta_{ij}),\label{eq:angle}\\
\dot{\psi}_{ij}=\omega_{0}\Phi_{ij}\cos(\omega_{0}\tau+\beta_{ij}),\label{eq:velocity}\\
\ddot{\psi}_{ij}=-\omega_{0}^{2}\Phi_{ij}\sin(\omega_{0}\tau+\beta_{ij}),\label{eq:acceleration}
\end{align}
where $\beta_{ij}\ (i,j\ =1,2)$ are phase differences between the pendula.

Equation (\ref{eq:2:a}) allows an estimation of the resultant force with
which the pendula act on the beam: 
\begin{equation}
F=-\sum_{i=1}^{2}\textbf{A}_{i1}(\ddot{\psi}_{i1}\cos\psi_{i1}-\dot{\psi}_{i1}^{2}\sin\psi_{i1})-\sum_{i=1}^{2}\textbf{A}_{i2}(\ddot{\psi}_{i2}\cos\psi_{i2}-\dot{\psi}_{i2}^{2}\sin\psi_{i2}).\label{eq:beam-1}
\end{equation}
Substituting Eqs (\ref{eq:angle}-\ref{eq:acceleration}) into Eq.
(\ref{eq:beam-1}) and considering the relation $\cos^{2}\alpha\sin\alpha=0.25\sin\alpha+0.25\sin3\alpha$,
one obtains: 
\begin{eqnarray}
F=\textbf{A}_{11}[\omega_{0}^{2}\Phi_{11}(1+0.25\Phi_{11}^{2})\sin(\omega_{0}\tau+\beta_{11})+\omega_{0}^{2}\Phi_{11}^{3}0.25\sin(3\omega_{0}\tau+3\beta_{11})]\nonumber \\
+\textbf{A}_{12}[\omega_{0}^{2}\Phi_{12}(1+0.25\Phi_{12}^{2})\sin(\omega_{0}\tau+\beta_{12})+\omega_{0}^{2}\Phi_{12}^{3}0.25\sin(3\omega_{0}\tau+3\beta_{12})]\nonumber \\
+\textbf{A}_{21}[\omega_{0}^{2}\Phi_{21}(1+0.25\Phi_{21}^{2})\sin(\omega_{0}\tau+\beta_{21})+\omega_{0}^{2}\Phi_{21}^{3}0.25\sin(3\omega_{0}\tau+3\beta_{21})]\nonumber \\
+\textbf{A}_{22}[\omega_{0}^{2}\Phi_{22}(1+0.25\Phi_{22}^{2})\sin(\omega_{0}\tau+\beta_{22})+\omega_{0}^{2}\Phi_{22}^{3}0.25\sin(3\omega_{0}\tau+3\beta_{22})].\label{eq:Force_dim}
\end{eqnarray}
Equation (\ref{eq:Force_dim}) is the right-hand side of equation of the beam
motion (\ref{eq:2:a}), hence we have:

\begin{equation}
\ddot{y}_{b}+\mathbf{K}y_{b}+\textbf{C}\dot{y}_{b}=F.
\end{equation}
Assuming that the damping coefficient $\textbf{C}$ is small, one gets:
\begin{eqnarray}
y_{b}=\sum_{i=1}^{2}\sum_{j=1}^{2}\textbf{X}_{1ij}\textbf{A}_{ij}\sin(\omega_{0}\tau+\beta_{ij})+\sum_{i=1}^{2}\sum_{j=1}^{2}\textbf{X}_{3ij}\textbf{A}_{ij}\sin(3\omega_{0}\tau+3\beta_{ij}),\nonumber \\
\ddot{y}_{b}=\sum_{i=1}^{2}\sum_{j=1}^{2}\textbf{A}_{1ij}\textbf{A}_{ij}\sin(\omega_{0}\tau+\beta_{ij})+\sum_{i=1}^{2}\sum_{j=1}^{2}9\textbf{A}_{3ij}\textbf{A}_{ij}\sin(3\omega_{0}\tau+3\beta_{ij}),\label{eq:BeamAccereration}
\end{eqnarray}
where: 
\begin{align}
X_{1ij}=\frac{\omega_{0}^{2}\Phi_{ij}(1+0.25\Phi_{ij}^{2})}{\textbf{K}-\omega_{0}^{2}},\quad X_{3ij}=\frac{0.25\omega_{0}^{2}\Phi_{ij}^{3}}{\textbf{K}-9\omega_{0}^{2}},\nonumber \\
\textbf{A}_{1ij}=\frac{-\omega_{0}^{4}\Phi_{ij}(1+0.25\Phi_{ij}^{2})}{\textbf{K}-\omega_{0}^{2}},\quad\textbf{A}_{3ij}=\frac{-0.25\omega_{0}^{4}\Phi_{ij}^{3}}{\textbf{K}-9\omega_{0}^{2}}.
\end{align}
Equations (\ref{eq:BeamAccereration}) represent the displacement and the
acceleration of the beam $M$, respectively.

\subsection{Energy balance of the system}

Multiplying Eq. (\ref{eq:2:a}) by the velocity of the beam $\dot{y}_{b}$,
we obtain: 
\begin{equation}
\ddot{y}_{b}\dot{y}_{b}+\mathbf{K}y_{b}\dot{y}_{b}=-\textbf{C}\dot{y}_{b}^{2}-\sum_{i=1}^{2}\textbf{A}_{i1}(\ddot{\psi}_{i1}\dot{y}_{b}\cos\psi_{i1}-\dot{\psi}_{i1}^{2}\dot{y}_{b}\sin\psi_{i1})-\sum_{i=1}^{2}\textbf{A}_{i2}(\ddot{\psi}_{i2}\dot{y}_{b}\cos\psi_{i2}-\dot{\psi}_{i2}^{2}\dot{y}_{b}\sin\psi_{i2}).\label{eq:EnergyBeam}
\end{equation}
Assuming that the motion of the pendulum is periodic with the period $T$
($T=2\pi/\omega_{0}$) and integrating Eq. (\ref{eq:EnergyBeam}),
we obtain the following energy balance: 
\begin{flalign}
\intop_{0}^{T}\ddot{y}_{b}\dot{y}_{b}d\tau+\intop_{0}^{T}\mathbf{K}y_{b}\dot{y}_{b}d\tau & =-\intop_{0}^{T}\textbf{C}\dot{y}_{b}^{2}d\tau-\intop_{0}^{T}\sum_{j=1}^{2}\left(\sum_{i=1}^{2}\textbf{A}_{ij}(\ddot{\psi}_{ij}\cos\psi_{ij}-\dot{\psi}_{ij}^{2}\sin\psi_{ij})\right)\dot{y}_{b}d\tau.\label{eq:LHS:energy_beam}
\end{flalign}
The left-hand side of Eq. (\ref{eq:LHS:energy_beam}) represents an increase
in the total energy of the beam which for the periodic oscillations
is equal to zero: 
\begin{equation}
\intop_{0}^{T}\ddot{y}_{b}\dot{y}_{b}d\tau+\intop_{0}^{T}\mathbf{K}y_{b}\dot{y}_{b}d\tau=0.
\end{equation}
The first component of the right-hand side of Eq. \eqref{eq:LHS:energy_beam}
represents the energy dissipated by the linear damper \textbf{C}: 
\begin{equation}
W_{beam}^{DAMP}=\intop_{0}^{T}\textbf{C}\dot{y}_{b}^{2}d\tau,\label{eq:BeamDamp}
\end{equation}
whereas the second component represents the work performed by horizontal components
of the force with which the double pendula act on the beam causing
its motion: 
\begin{equation}
W_{beam}^{DRIVE}=-\intop_{0}^{T}\sum_{j=1}^{2}\left(\sum_{i=1}^{2}\textbf{A}_{ij}(\ddot{\psi}_{ij}\cos\psi_{ij}-\dot{\psi}_{ij}^{2}\sin\psi_{ij})\right)\dot{y}_{b}d\tau.\label{eq:BeamDrive}
\end{equation}
Substituting Eqs (\ref{eq:BeamDamp}) and (\ref{eq:BeamDrive}) into
Eq. (\ref{eq:LHS:energy_beam}), we get: 
\begin{equation}
W_{beam}^{DRIVE}-W_{beam}^{DAMP}=0.
\end{equation}
Multiplying the equation of the upper pendulum (Eq. (\ref{eq:2:b}))
by the velocity $\dot{\psi}_{i1}$, we obtain: 
\begin{eqnarray}
\delta_{i1}\ddot{y}_{b}\dot{\psi}_{i1}\cos\psi_{i1}+\textbf{L}_{i1}\ddot{\psi}_{i1}\dot{\psi}_{i1}+\textbf{L}_{i3}\ddot{\psi}_{i2}\dot{\psi}_{i1}\cos(\psi_{i1}-\psi_{i2}) & = & -\textbf{L}_{i3}\dot{\psi}_{i1}\dot{\psi}_{i2}^{2}\sin(\psi_{i1}-\psi_{i2})\nonumber \\
-\textbf{G}_{i1}\dot{\psi}_{i1}\sin(\psi_{i1})-\mathbf{C}_{vdp}(1-\mu\psi_{i1}^{2})\dot{\psi}_{i1}^{2}+\mathbf{C}_{i2}(\dot{\psi}_{i2}-\dot{\psi}_{i1})\dot{\psi}_{i1}.\label{eq:EnergyPendG}
\end{eqnarray}
Assuming that the oscillations of the pendula are periodic with the period
$T$ and integrating Eq. (\ref{eq:EnergyPendG}), one obtains the following
energy balance: 
\begin{gather}
\intop_{0}^{T}\textbf{L}_{i1}\ddot{\psi}_{i1}\dot{\psi}_{i1}d\tau+\intop_{0}^{T}\textbf{G}_{i1}\dot{\psi}_{i1}\sin\psi_{i1}d\tau=-\intop_{0}^{T}\delta_{i1}\ddot{y}_{b}\dot{\psi}_{i1}\cos\psi_{i1}d\tau\nonumber \\
-\intop_{0}^{T}\textbf{L}_{i3}(\dot{\psi}_{i1}\dot{\psi}_{i2}^{2}\sin(\psi_{i1}-\psi_{i2})+\ddot{\psi}_{i2}\dot{\psi}_{i1}\cos(\psi_{i1}-\psi_{i2}))d\tau-\nonumber \\
\intop_{0}^{T}\mathbf{C}_{vdp}\dot{\psi}_{i1}^{2}d\tau+\intop_{0}^{T}\mathbf{C}_{vdp}\mu\psi_{i1}^{2}\dot{\psi}_{i1}^{2}d\tau+\intop_{0}^{T}\mathbf{C}_{i2}\dot{\psi}_{i2}\dot{\psi}_{i1}d\tau-\intop_{0}^{T}\mathbf{C}_{i2}\dot{\psi}_{i1}^{2}d\tau.\label{eq:EnergySum}
\end{gather}
The left side of Eq. (\ref{eq:EnergySum}) represents the total energy
of the upper pendula, which in the case of periodic oscillations is equal
to zero: 
\begin{equation}
\intop_{0}^{T}\textbf{L}_{i1}\ddot{\psi}_{i1}\dot{\psi}_{i1}d\tau+\intop_{0}^{T}\textbf{G}_{i1}\dot{\psi}_{i1}\sin\psi_{i1}d\tau=0.
\end{equation}
The first component of the right side of Eq. (\ref{eq:EnergySum})
represents the energy which is transferred to the beam: 
\begin{equation}
W_{i1}^{SYN}=\intop_{0}^{T}\delta_{i1}\ddot{y}_{b}\dot{\psi}_{i1}\cos\psi_{i1}d\tau.\label{eq:EnergyToBeamU}
\end{equation}
The second component describes the energy which is transferred to the
lower pendulum: 
\begin{equation}
W_{i1}^{SYN~P}=-\intop_{0}^{T}\textbf{L}_{i3}(\dot{\psi}_{i1}\dot{\psi}_{i2}^{2}\sin(\psi_{i1}-\psi_{i2})+\ddot{\psi}_{i2}\dot{\psi}_{i1}\cos(\psi_{i1}-\psi_{i2}))d\tau,\label{eq:EnergyToLowerPen}
\end{equation}
and the third component describes the energy which is supplied to
the system by the van der Pol damper in one-period oscillations: 
\begin{equation}
W_{i1}^{DAMP}=-\intop_{0}^{T}(\mathbf{C}_{vdp}+\mathbf{C}_{i2})\dot{\psi}_{i1}^{2}-\mathbf{C}_{i2}\dot{\psi}_{i2}\dot{\psi_{i1}}d\tau.\label{eq:EnergyDrive}
\end{equation}
Finally, the last component represents the energy dissipated by the van
der Pol damper: 
\begin{equation}
W_{i1}^{SELF}=-\int_{0}^{T}\mu\mathbf{C}_{vdp}\psi_{i1}^{2}\dot{\psi}_{i1}^{2}d\tau,\label{eq:EnergyDamped}
\end{equation}
Substituting Eqs (\ref{eq:EnergyToBeamU} - \ref{eq:EnergyDamped})
into Eq. (\ref{eq:EnergySum}), we obtain the following relation: 
\[
W_{i1}^{SYN~P}-W_{i1}^{SYN}+W_{i1}^{SELF}+W_{i1}^{DAMP}=0,
\]
Multiplying the equation of the lower pendulum (Eq. (\ref{eq:2:c}))
by the velocity $\dot{\psi}_{i2}$, one gets: 
\begin{gather}
\delta_{i2}\ddot{y}_{b}\dot{\psi}_{i2}\cos\psi_{i2}+\textbf{L}_{i3}\ddot{\psi}_{i1}\dot{\psi}_{i2}\cos(\psi_{i1}-\psi_{i2})+\textbf{L}_{i2}\dot{\psi}_{i2}\ddot{\psi}_{i2}=\label{eq:EnergyLP}\\
\textbf{L}_{i3}\dot{\psi}_{i1}^{2}\dot{\psi}_{i2}\sin(\psi_{i1}-\psi_{i2})-\textbf{G}_{i2}\dot{\psi}_{i2}\sin(\psi_{i2})-\mathbf{C}_{i2}(\dot{\psi}_{i2}-\dot{\psi}_{i1})\dot{\psi}_{i2}.\nonumber 
\end{gather}
Assuming that the oscillations of the pendulum are periodic with the period
$T$, the integration of Eq. (\ref{eq:EnergyLP}) gives the following
energy balance: 
\begin{gather}
\intop_{0}^{T}\textbf{L}_{i2}\dot{\psi}_{i2}\ddot{\psi}_{i2}d\tau+\intop_{0}^{T}\textbf{G}_{i2}\dot{\psi}_{i2}\sin(\psi_{i2})d\tau=-\intop_{0}^{T}\beta_{i2}\ddot{y}_{b}\dot{\psi}_{i2}\cos\psi_{i2}d\tau-\label{eq:EnergyLowerPendulum-1}\\
\intop_{0}^{T}\textbf{L}_{i3}(\dot{\psi}_{i1}^{2}\dot{\psi}_{i2}\sin(\psi_{i1}-\psi_{i2})-\ddot{\psi}_{i1}\dot{\psi}_{i2}\cos(\psi_{i1}-\psi_{i2}))d\tau-\intop_{0}^{T}\mathbf{C}_{i2}\dot{\psi}_{i2}^{2}d\tau+\intop_{0}^{T}\mathbf{C}_{i2}\dot{\psi}_{i1}\dot{\psi}_{i2}d\tau.\nonumber 
\end{gather}
The left side of Eq. (\ref{eq:EnergyLowerPendulum-1}) represents
the total energy of the lower pendulum, which in the case of periodic oscillations
is equal to zero: 
\begin{equation}
\intop_{0}^{T}\textbf{L}_{i2}\dot{\psi}_{i2}\ddot{\psi}_{i2}d\tau+\intop_{0}^{T}\textbf{G}_{i2}\dot{\psi}_{i2}\sin(\psi_{i2})d\tau=0.
\end{equation}
The first component of the right side of Eq. (\ref{eq:EnergyLowerPendulum-1})
represents the energy which is transferred to the beam via the upper pendulum
or to the next pendulum via the upper pendulum and the beam: 
\begin{equation}
W_{i2}^{SYN}=\intop_{0}^{T}\delta_{i2}\ddot{y}_{b}\dot{\psi}_{i2}\cos\psi_{i2}d\tau.\label{eq:5}
\end{equation}
The second component describes the energy which is transferred to
the upper pendulum: 
\begin{equation}
W_{i2}^{SYN~P}=-\intop_{0}^{T}\textbf{L}_{i3}(\dot{\psi}_{i1}^{2}\dot{\psi}_{i2}\sin(\psi_{i1}-\psi_{i2})-\ddot{\psi}_{i1}\dot{\psi}_{i2}\cos(\psi_{i1}-\psi_{i2}))d\tau.\label{eq:6}
\end{equation}
and the last component represents the energy dissipated by the damper:
\begin{equation}
W_{i2}^{DAMP}=-\intop_{0}^{T}\mathbf{C}_{i2}(\dot{\psi}_{i2}-\dot{\psi_{i1}})\dot{\psi_{i2}}d\tau\label{eq:7}
\end{equation}
Substituting Eqs (\ref{eq:5} - \ref{eq:7}) into Eq. (\ref{eq:EnergyLowerPendulum-1}),
one obtains the following relation: 
\[
W_{i2}^{SYN~P}-W_{i2}^{SYN}+W_{i2}^{DAMP}=0.
\]

\subsection{Energy transfer between the upper and lower pendula}

The energy transferred from the upper to lower pendulum is given
by: 
\begin{equation}
W_{i1}^{SYN~P}=-\intop_{0}^{T}\textbf{L}_{i3}(\ddot{\psi}_{i2}\cos(\psi_{i1}-\psi_{i2})+\dot{\psi}_{i2}^{2}\sin(\psi_{i1}-\psi_{i2}))\dot{\psi}_{i1}d\tau,\label{eq:PenUtoL}
\end{equation}
and the energy transferred from the lower to upper pendulum is:
\begin{equation}
W_{i2}^{SYN~P}=-\intop_{0}^{T}\textbf{L}_{i3}(\ddot{\psi}_{i1}\cos(\psi_{i1}-\psi_{i2})-\dot{\psi}_{i1}^{2}\sin(\psi_{i1}-\psi_{i2}))\dot{\psi}_{i2}d\tau.\label{eq:PenLtoU}
\end{equation}
Taking into account Eqs (\ref{eq:angle} - \ref{eq:acceleration}),
Eq. (\ref{eq:PenLtoU}) takes the form: 
\begin{gather}
W_{i1}^{SYN~P}=-\textbf{L}_{i3}\intop_{0}^{T}(-\omega_{0}^{2}\Phi_{i2}\sin(\omega_{0}t+\beta_{i2})\cos(\Phi_{i1}\sin(\omega_{0}t+\beta_{i1})-\Phi_{i2}\sin(\omega_{0}t+\beta_{i2}))\nonumber \\
+\omega_{0}^{2}\Phi_{i2}^{2}\cos^{2}(\omega_{0}t+\beta_{i2})\sin(\Phi_{i1}\sin(\omega_{0}t+\beta_{i1})-\Phi_{i2}\sin(\omega_{0}t+\beta_{i2})))\omega_{0}\Phi_{i1}\cos(\omega_{0}t+\beta_{i1})d\tau\\
=\textbf{L}_{i3}\pi\omega_{0}^{2}\Phi_{i1}\Phi_{i2}\sin(\beta_{i2}-\beta_{i1}),\nonumber 
\end{gather}
and 
\begin{eqnarray}
W_{i2}^{SYN~P}=\textbf{L}_{i3}\pi\omega_{0}^{2}\Phi_{i1}\Phi_{i2}\sin(\beta_{i1}-\beta_{i2})=-W_{i1}^{SYN~P},
\end{eqnarray}
The synchronization between the lower and upper pendula occurs
when: 
\begin{equation}
W_{i1}^{SYN~P}=0\quad\Rightarrow\quad\sin(\beta_{i1}-\beta_{i2})=0.\label{eq:WarSin}
\end{equation}
Condition (\ref{eq:WarSin}) is fulfilled when: 
\begin{equation}
\beta_{i1}=\beta_{i2}\vee(\beta_{i1}=0\wedge\beta_{i2}=\pi).\label{eq:cond_synch}
\end{equation}
In the first case, the oscillations of the upper and lower pendula
are in-phase, i.e., the pendula move in the same directions, whereas in the
second case they are in anti-phase, i.e., the pendula move in the opposite directions.
For low oscillations, limit conditions (\ref{eq:cond_synch})
define two normal modes of oscillations \cite{ref1}.

\subsection{Synchronization of the double pendula}

In each equation of the pendulum motion, there is a component 
influencing the beam motion 
\begin{equation}
M_{ij}^{SYN}=\delta_{ij}\ddot{y}_{b}\cos\psi_{ij},
\end{equation}
which is called the synchronization momentum (torque). The work done
by this momentum during one period is equal to zero. 
\begin{equation}
W_{ij}^{SYN}=\int_{0}^{T}\delta_{ij}\ddot{y}_{b}\cos\psi_{ij}\dot{\psi}_{ij}d\tau=0.\label{eq:PendulumToBeam}
\end{equation}
Substituting Eqs (\ref{eq:angle}, \ref{eq:velocity}) and (\ref{eq:BeamAccereration})
into (\ref{eq:PendulumToBeam}) and performing the linearization, we
arrive at: 
\begin{align}
W_{11}^{SYN}= & \int_{0}^{T}\delta_{11}\left(\sum_{i=1}^{2}\sum_{j=1}^{2}\textbf{A}_{1ij}\textbf{A}_{ij}\sin(\omega_{0}t+\beta_{ij})+\sum_{i=1}^{2}\sum_{j=1}^{2}\textbf{A}_{3ij}\textbf{A}_{ij}\sin(3\omega_{0}t+3\beta_{ij})\right)\omega_{0}\Phi_{11}\cos(\omega_{0}t+\beta_{11})d\tau=\label{eq:BeamSynchronization}\\
= & \delta_{11}\omega_{0}\pi\Phi_{11}\sum_{i=1}^{2}\sum_{j=1}^{2}\textbf{A}_{1ij}\textbf{A}_{ij}\sin(\beta_{ij}-\beta_{11})=\nonumber \\
= & \xi\delta_{11}\Phi_{11}\left[\sum_{i=1}^{2}\sum_{j=1}^{2}\Theta_{ij}M_{ij}\sin(\beta_{ij}-\beta_{11})\right]=0,\nonumber \\
W_{12}^{SYN}= & \xi\delta_{12}\Phi_{12}\left[\sum_{i=1}^{2}\sum_{j=1}^{2}\Theta_{ij}M_{ij}\sin(\beta_{ij}-\beta_{12})\right]=0,\nonumber \\
W_{21}^{SYN}= & \xi\delta_{21}\Phi_{21}\left[\sum_{i=1}^{2}\sum_{j=1}^{2}\Theta_{ij}M_{ij}\sin(\beta_{ij}-\beta_{21})\right]=0,\nonumber \\
W_{22}^{SYN}= & \xi\delta_{22}\Phi_{22}\left[\sum_{i=1}^{2}\sum_{j=1}^{2}\Theta_{ij}M_{ij}\sin(\beta_{ij}-\beta_{22})\right]=0,\nonumber 
\end{align}
where: 
\begin{align}
\xi=\frac{-\omega_{0}^{5}\pi}{{M}l_{b}(\textbf{K}-\omega_{0}^{2})},\quad M_{i1}=(m_{i1}+m_{i2})l_{i1},\quad M_{i2}=m_{i2}l_{i2},\quad\Theta_{ij}=\Phi_{ij}(1+0.25\Phi_{ij}^{2}).
\end{align}
Equations (\ref{eq:BeamSynchronization}) allow the calculation of the phase
angles $\beta_{ij}$ for which the synchronization of periodic
oscillations of the pendula occurs. The synchronization occurs when the
following equations are fulfilled:

\begin{equation}
\begin{array}{c}
\Theta_{12}M_{12}\sin(\beta_{12}-\beta_{11})+\Theta_{21}M_{21}\sin(\beta_{21}-\beta_{11})+\Theta_{22}M_{22}\sin(\beta_{22}-\beta_{11})=0,\\
\Theta_{11}M_{11}\sin(\beta_{11}-\beta_{12})+\Theta_{21}M_{21}\sin(\beta_{21}-\beta_{12})+\Theta_{22}M_{22}\sin(\beta_{22}-\beta_{12})=0,\\
\Theta_{11}M_{11}\sin(\beta_{11}-\beta_{21})+\Theta_{12}M_{12}\sin(\beta_{12}-\beta_{21})+\Theta_{22}M_{22}\sin(\beta_{22}-\beta_{21})=0,\\
\Theta_{11}M_{11}\sin(\beta_{11}-\beta_{22})+\Theta_{12}M_{12}\sin(\beta_{12}-\beta_{22})+\Theta_{21}M_{21}\sin(\beta_{21}-\beta_{22})=0.
\end{array}\label{eq:SynchCondition}
\end{equation}

\begin{figure}[H]
\begin{centering}
\includegraphics{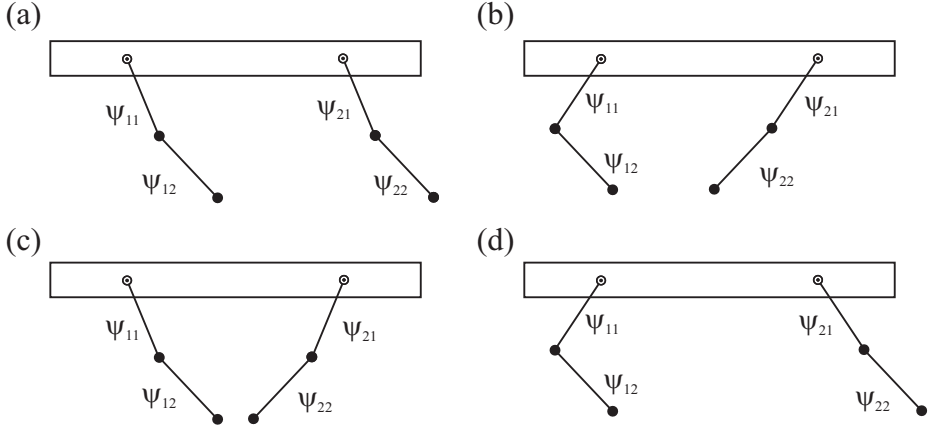}
\par\end{centering}

\caption{\label{fig:KindSynch}Synchronous states of the system (\ref{eq:DimForm}):
(a) upper and lower pendula in phase: $\psi_{11}=\psi_{21}$ and $\psi_{12}=\psi_{22}$,
(b) upper pendula in phase, lower pendula in anti-phase: $\psi_{11}=\psi_{21}$
and $\psi_{12}=-\psi_{22}$, (c) upper and lower pendula in anti-phase:
$\psi_{11}=-\psi_{21}$ and $\psi_{12}=-\psi_{22}$ (d) upper pendula
in anti-phase, lower pendula in phase: $\psi_{11}=-\psi_{21}$ and
$\psi_{12}=\psi_{22}$. }
\end{figure}

Equations (\ref{eq:SynchCondition}) are fulfilled for $\beta_{ij}$,
which are combinations of $0$ and $\pi$. Assuming that $\beta_{11}=0$,
one can identify the following pendulum configurations which are
presented in Fig. 2(a-d). The first type is the configuration shown
in Fig. \ref{fig:KindSynch}(a). Both the upper and lower pendula are
phase synchronized, i.e., $\psi_{11}=\psi_{21}$ and $\psi_{12}=\psi_{22}$
($\beta_{11}=\beta_{12}=\beta_{21}=\beta_{22}=0$ or $\beta_{11}=\beta_{21}=0,\ \beta_{12}=\beta_{22}=\pi$).
The upper and lower pendula are synchronized in phase and anti-phase, respectively,
i.e., $\psi_{11}=\psi_{21}$ and $\psi_{12}=-\psi_{22}$
in the configuration from Fig. \ref{fig:KindSynch}(b) ($\beta_{11}=\beta_{12}=\beta_{21}=0$,
$\beta_{22}=\pi$ or $\beta_{11}=\beta_{12}=\beta_{22}=0$, $\beta_{21}=\pi$).
Figure \ref{fig:KindSynch}(c) presents the case when both the upper and
lower pendula are synchronized in anti-phase, i.e., $\psi_{11}=-\psi_{21}$
and $\psi_{12}=-\psi_{22}$ ($\beta_{11}=\beta_{12}=0$, $\beta_{21}=\beta_{22}=\pi$
or $\beta_{11}=\beta_{22}=0$, $\beta_{12}=\beta_{21}=\pi$). Finally,
in Fig. \ref{fig:KindSynch}(d), we present the case when the upper pendula
are in anti-phase and the lower pendula are in phase $\psi_{11}=-\psi_{21}$
and $\psi_{12}=\psi_{22}$ ($\beta_{11}=\beta_{12}=\beta_{22}=0$,
$\beta_{21}=\pi$ or $\beta_{11}=0$, $\beta_{21}=\beta_{12}=\beta_{22}=\pi$).

\section{Numerical investigations\label{sec:Numerical-results}}

In our numerical calculations, we use the Auto 07p \cite{refAuto}
continuation toolbox to obtain periodic solutions. To start path-following,
we integrate Eqs (\ref{eq:2:a}-\ref{eq:2:c}) with the fourth-order Runge-Kutta
method. We consider the following parameter values: $m_{11}=m_{12}=m_{21}=m_{22}=1.0\:[\mathrm{kg}]$,
$M=10.0\:[\mathrm{kg}]$, $l_{11}=l_{12}=l_{21}=l_{22}=0.2485\:[\mathrm{m}]$,
$k_{x}=4.0\:[\mathrm{N/m}]$, $c_{x}=1.53\:[\mathrm{Ns/m}]$, $c_{vdp}=-0.1\:[\mathrm{Ns/m}]$,
$\mu=60.0\:[\mathrm{m^{-2}}]$, $c_{i2}=0.0016\:[\mathrm{Ns/m}]$,
which yield the following dimensionless coefficients $\textbf{A}_{i1}=0.0354986$,
$\textbf{A}_{i2}=0.01774933$, $\delta_{i1}=0.142857$, $\delta_{i2}=0.0714286$,
$\textbf{L}_{i1}=0.035986$, $\textbf{L}_{i2}=0.0177493$, $\textbf{L}_{i3}=0.0177493$,
$\textbf{G}_{i1}=0.0354986$, $\textbf{G}_{i2}=0.0177493$, $\textbf{C}_{vdp}=-0.00457491$,
$\textbf{C}_{i2}=0.0000714286$, $\textbf{C}=0.0173934$, $\textbf{K}=0.00723723$.
Our bifurcation parameters are masses of the pendula and the beam. To hold
an intuitive physical interpretation, we change dimensional masses, but all the calculation are performed for
dimensionless equations.

\subsection{Periodic solutions to the pendula with identical masses}

\begin{figure}[h]
\begin{centering}
\includegraphics[scale=0.8]{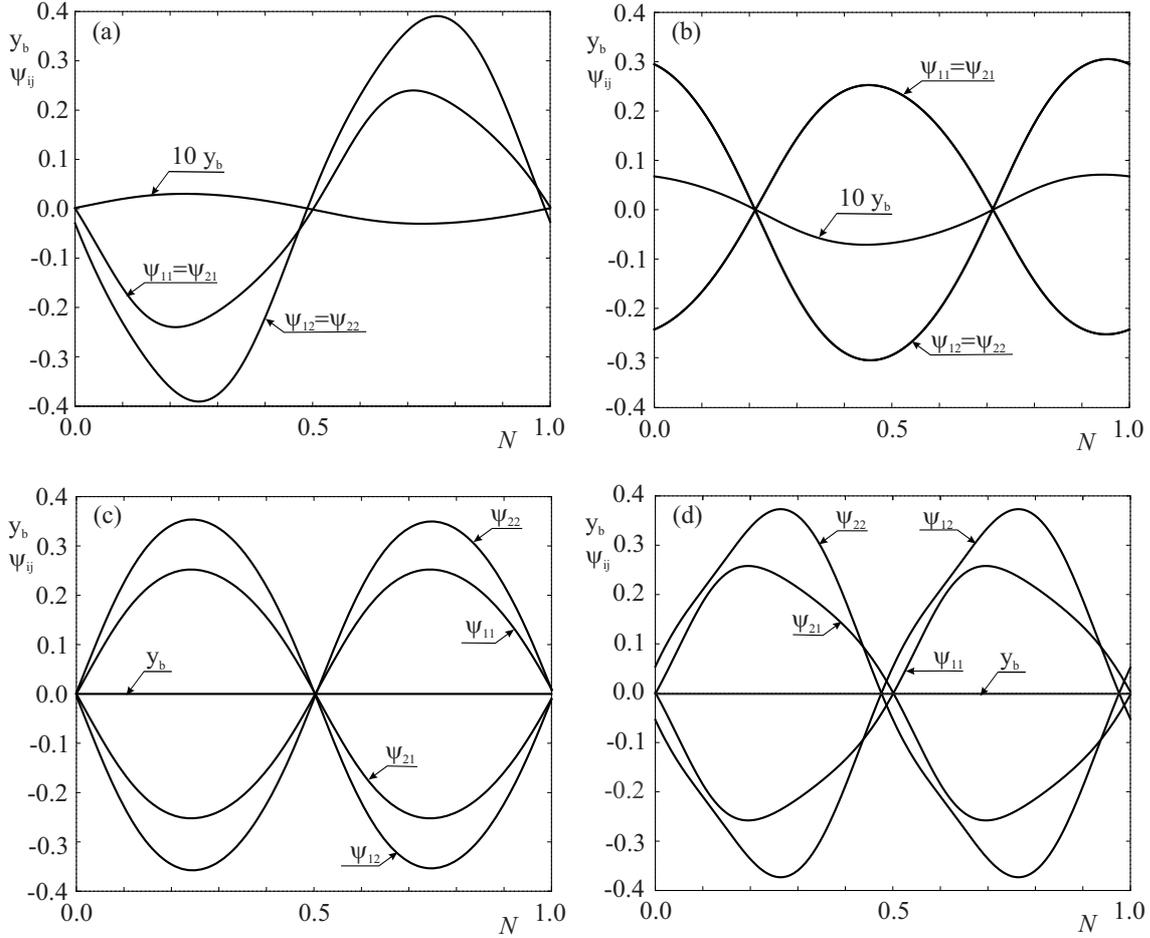}
\par\end{centering}

{\scriptsize{\caption{Pendulum and beam displacements for one period of motion ($N=1$)
for four different periodic solutions in the case of identical masses
of the pendula: $m_{11}=m_{12}=m_{21}=m_{22}=1.0\:[\mathrm{kg}]$. The
displacement $y_{b}$ of the beam $M$ is shown $10$ times magnified.
(a) pendulum configuration from Fig. 2(a) with the period $T=7.233$,
(b) pendulum configuration from Fig. 2(a) with the period $T=3.362$,
(c) pendulum configuration from Fig. 2(c) with the period
$T=3.748$, and (d) pendulum configuration from Fig. 2(c) with the period
$T=8.266$.\label{fig:Idem_PS} }
}}
\end{figure}

Depending on the initial conditions, we observe four different synchronous
states of system (\ref{eq:2:a}-\ref{eq:2:c}) as shown in Fig.
\ref{fig:Idem_PS}(a-d). The in-phase motion is represented by two
periodic solutions: the first type is characterized by a lack of
phase differences in the pendulum angular positions: $\beta_{11}=\beta_{21}=\beta_{12}=\beta_{22}$
(see Fig. \ref{fig:Idem_PS}(a)), whereas the second one - by a phase difference
between the upper and lower pendula in each double pendulum: $\beta_{11}=\beta_{21}$,
$\beta_{12}=\beta_{22}$ and $\beta_{i1}-\beta_{i2}=\pi$, $i=1,2$
(see Fig. \ref{fig:Idem_PS}(b)). In both cases, the displacements
of the upper and lower pendula of each double pendulum are identical,
i.e., $\psi_{11}=\psi_{21}$, $\psi_{12}=\psi_{22}$. The beam
motion is in anti-phase to the upper pendula and in-phase (Fig. \ref{fig:Idem_PS}(a))
or anti-phase (Fig. \ref{fig:Idem_PS}(b)) to the lower ones. These two
configurations correspond to the analytically predicted synchronous
state presented in Fig. \ref{fig:KindSynch}(a). The second type,
the anti-phase motion, for
which the beam is not moving, is shown in Fig. \ref{fig:Idem_PS}(c,d). We can also distinguish two types of
this periodic solution, both characterized by the following phase
differences of the pendulum displacements: $\beta_{11}-\beta_{21}=\pi$
and $\beta_{12}-\beta_{22}=\pi$, but different phase shifts between
the pendula in each double pendulum: $\beta_{i1}-\beta_{i2}=\pi$, $i=1,2$
(see Fig. \ref{fig:Idem_PS}(c)) and $\beta_{i1}-\beta_{i2}=0$, $i=1,2$
(see Fig. \ref{fig:Idem_PS}(d)). The beam $M$ is at rest, because
reaction forces acting on the beam are vanishing. These pendulum
configurations correspond to the theoretically predicted synchronous state
presented in Fig. \ref{fig:KindSynch}(c). Note that for this type of the synchronous
state, amplitudes of pendulum oscillations can be estimated analytically.
Substituting $\beta_{11}=\beta_{12}=0$ and $\beta_{21}=\beta_{22}=\pi$
in Eqs (\ref{eq:SynchCondition}), one can derive an analytical formula
for amplitudes of pendulum oscillations. The amplitudes of the
upper pendula can be approximated by: 
\begin{align}
\Phi_{11} & =\Phi_{21}=2\sqrt{\frac{1}{\mu}}.\label{eq:amplitudeUpperPend}
\end{align}
The approximate values of the amplitudes of oscillations of the lower
pendula can be calculated from the following condition:

\begin{align}
\frac{1}{4}\omega_{0}\pi(4\textbf{C}_{i2}(\Phi_{i1}^{2}+\Phi_{i2}^{2})+\textbf{C}_{vdp}\Phi_{i1}^{2}(4-\mu\Phi_{i1}^{2})-8\textbf{C}_{i2}\Phi_{i1}\Phi_{i2}\cos(\beta_{i1}-\beta_{i2})+\label{eq:amplitudeLowerPend}\\
\Phi_{i1}\Phi_{i2}(\Phi_{i2}^{2}-\Phi_{i1}^{2})\textbf{L}_{i3}\omega_{0}\sin(\beta_{i1}-\beta_{i2})) & =0\nonumber 
\end{align}
For $i=1,2$, formulae (\ref{eq:amplitudeUpperPend}) and (\ref{eq:amplitudeLowerPend})
give good approximation of the numerical values, e.g., for the parameter
values in Fig. \ref{fig:Idem_PS}(c,d), the analytically calculated
amplitudes of the upper and lower pendula are $\Phi_{11}=\Phi_{21}=0.2581$
and $\Phi_{12}=\Phi_{22}=0.4244$, respectively, whereas the numerical values are $\Phi_{11}=\Phi_{21}=0.2522$
and $\Phi_{12}=\Phi_{22}=0.3536$ for Fig. \ref{fig:Idem_PS}(c) and
$\Phi_{12}=\Phi_{22}=0.2579$ and $\Phi_{12}=\Phi_{22}=0.3732$ for
Fig. \ref{fig:Idem_PS}(d). In Fig. \ref{fig:Idem_PS}(a,d), one can
see that the pendula do not pass through zero (the hanging down position)
at the same moment of time, whereas in Fig. \ref{fig:Idem_PS}(b,c)
the pendula cross this position simultaneously. The phase shift is observed
only when the lower and upper pendula in each double pendulum are oscillating
in-phase with non-zero damping between them. 

\noindent We do not observe the configurations shown in Fig. 2(b,d) because
each of the double pendulum has to reach different normal
modes of oscillations for the same frequency for both of them. This is proven to be
impossible in the low oscillation approximation \cite{ref1} (the
angular positions have to be much higher than the one considered in this paper).

\begin{figure}
\begin{centering}
\includegraphics[scale=0.8]{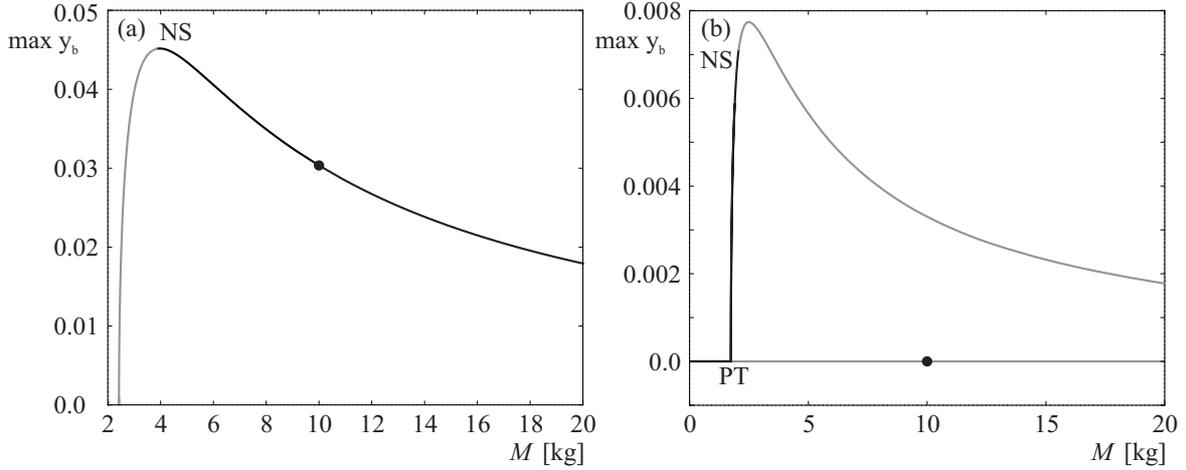}
\par\end{centering}

\caption{One-parameter path-following of periodic solutions for the varying mass
$M$ of the beam: (a) in-phase motion from Fig. \ref{fig:Idem_PS}(a)
and (b) anti-phase motion from Fig. \ref{fig:Idem_PS}(c). The black
and gray lines correspond to stable and unstable periodic solutions,
respectively. The abbreviation NS stands for the Neimark-Sacker bifurcation and PT denotes the pitchfork
bifurcation. Bifurcations along unstable branches are neglected.
The starting points of continuation are marked by black dots. \label{fig:One_par} }
\end{figure}

The periodic solutions presented in Fig. \ref{fig:Idem_PS}(a,b,d)
are stable, whereas the one from Fig. \ref{fig:Idem_PS}(c) is unstable.
To show how a change in the natural frequency of the beam affects
the stability of the periodic solutions obtained, we calculate one-parameter
bifurcation diagrams. We choose the beam
mass $M$ as the bifurcation parameter and vary it in the range from $0.01\:[\mathrm{kg}]$ to $20.0\:[\mathrm{kg}]$.
In the case of two solutions: one in-phase (Fig. \ref{fig:Idem_PS}(b))
and one anti-phase (Fig. \ref{fig:Idem_PS}(d)), we do not observe
any destabilization of periodic solutions. For the two others, we present
bifurcation diagrams showing the maximum amplitudes
of the beam oscillation $\mathrm{max}\: y_{b}$ on the vertical axes. The black and gray
colors of branches correspond to stable and unstable periodic solutions.
For the branch presented in Fig. \ref{fig:One_par}(a), we start a continuation
from the in-phase periodic solution shown in Fig. \ref{fig:Idem_PS}(a).
Originally, the stable periodic orbit becomes unstable with a decreasing beam
mass $M$ in the Neimark-Sacker bifurcation for $M=3.88\:[\mathrm{kg}]$.
The Neimark-Sacker bifurcation point corresponds also to the maximum amplitude
of the beam. In Fig. \ref{fig:One_par}(b), we show a continuation
of the anti-phase oscillations of the pendula (Fig. \ref{fig:Idem_PS}(c)).
The stabilization of this type of the periodic solution occurs in the supercritical
pitchfork bifurcation (two new branches emerge) for $M=1.737\:[\mathrm{kg}]$.
In the symmetric anti-phase motion, the beam is at rest and the maximum
amplitudes of the pendula remain the same. For the asymmetric periodic
solutions along two overlapping branches, the beam is oscillating with
a low amplitude and we observe a difference between amplitudes of
the pendula. The asymmetric motion destabilizes with an increase in the beam
mass $M$ in Neimark-Sacker bifurcations for $M=2.06\:[\mathrm{kg}]$. 

\subsection{Periodic solutions of the pendula with different masses - exploring
symmetry}

In this subsection, we investigate the stability of symmetric motion
of the pendula from Fig. \ref{fig:Idem_PS}(c,d), which
corresponds to anti-phase synchronization states. We decrease the masses
$m_{12}$ and $m_{22}$ of the lower pendula in the range $\left(0.0,\:1.0\right]\:[\mathrm{kg}]$.
We choose masses of the lower pendula as the bifurcation parameter because
we want to avoid a situation where the lightweight upper pendula excite the
much heavier lower pendula. 

\begin{figure}
\begin{centering}
\includegraphics[scale=0.8]{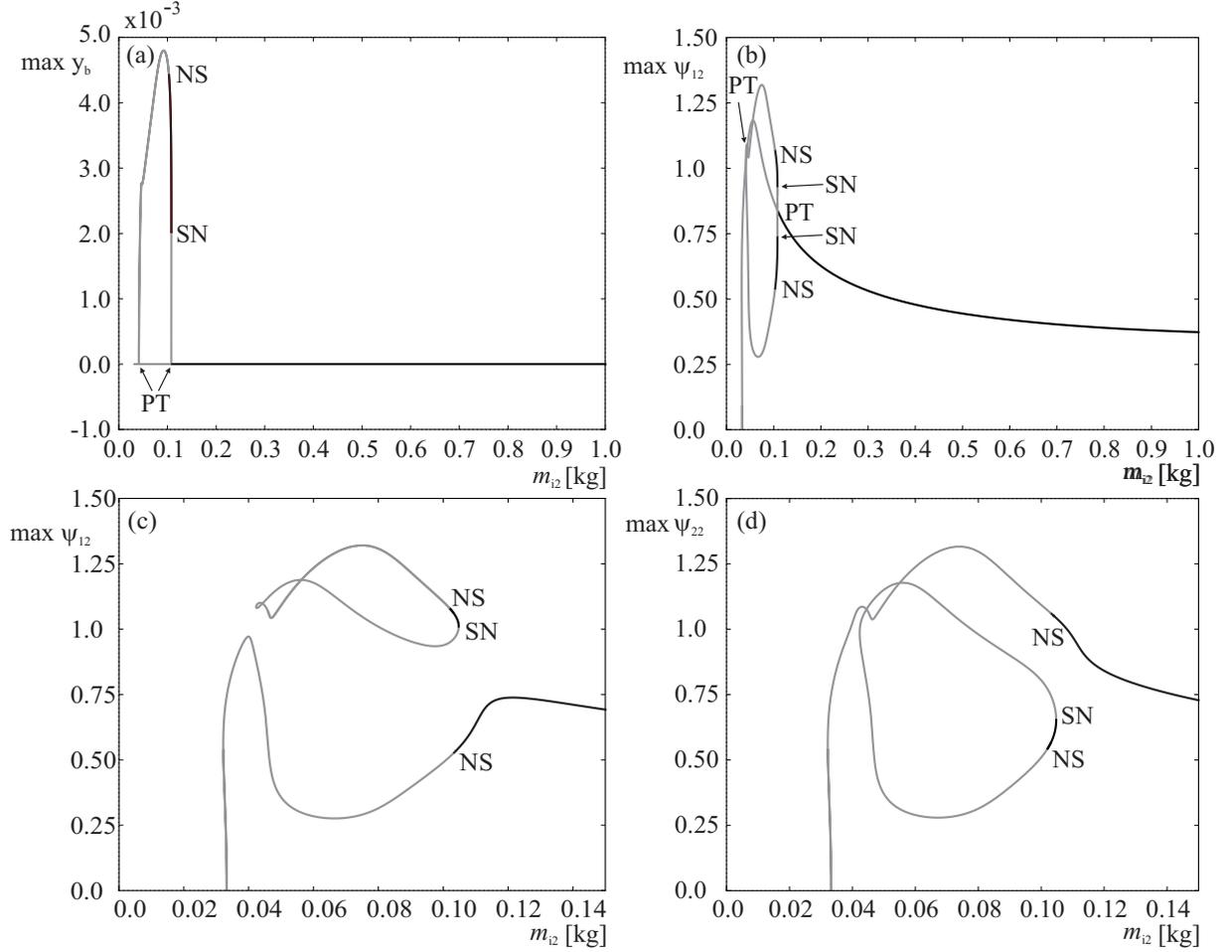}
\par\end{centering}

\caption{In (a,b) we show one-parameter path-following of the anti-phase synchronous
motion starting from the periodic solution shown in Fig. \ref{fig:Idem_PS}(d).
The changes in the  maximum amplitude of the beam $\mathrm{max}\: y_{b}$ (a)
and the second upper pendulum $\mathrm{max}\:\psi_{12}$ (b) are shown
for decreasing masses of the lower pendula $m_{i2}=\left(0.0,\:1.0\right]\:[\mathrm{kg}]$
($i=1,2$), whereas the upper pendula have masses equal to $1.0\:[\mathrm{kg}]$.
In (c,d) the same calculations are performed for asymmetrical masses
of the upper pendula ($m_{11}=1.0\:[\mathrm{kg}]$ , $m_{21}=0.99\:[\mathrm{kg}]$)
in the range $m_{i2}\in\left(0.0,\:0.15\right)\:[\mathrm{kg}]$ ($i=1,2$),
for $m_{i2}\in\left(0.15,\:1.0\right)\:[\mathrm{kg}]$ ($i=1,2$),
solutions are stable. The black and gray lines correspond to stable
and unstable periodic solutions, respectively. The abbreviations correspond
to: PT - pitchfork bifurcation, NS - Neimark-Saker bifurcation, and
SN - saddle-node bifurcation. Bifurcations along unstable branches
are neglected.\label{fig:PP_anti_phase}}
\end{figure}

In Fig. \ref{fig:PP_anti_phase}(a,b) we present bifurcation diagrams,
i.e., the maximum amplitudes of the beam (a) and the second upper pendulum (b)
for decreasing masses of the lower pendula. As an initial state, we take
the anti-phase periodic solution for which all pendula have identical
masses (see Fig. \ref{fig:Idem_PS}(d)). For $m_{12}=m_{22}=0.108\:[\mathrm{kg}]$,
the symmetry is broken in the subcritical pitchfork bifurcation. We
observe an appearance of two unstable branches which stabilize in
saddle-node bifurcations for $m_{12}=m_{22}=0.107\:[\mathrm{kg}]$,
a further loss of stability occurs in the supercritical Neimark-Sacker
bifurcations for $m_{12}=m_{22}=0.105\:[\mathrm{kg}]$, hence two
stable quasi-periodic solutions appear. This scenario is observed only
when the system has symmetry. The bifurcation diagram for a system without
symmetry ($m_{11}=1.0\:[\mathrm{kg}]$ and $m_{12}=0.99\:[\mathrm{kg}]$)
is shown in Fig. \ref{fig:PP_anti_phase}(c,d). We present the maximum
amplitudes of the upper pendula in the range $m_{i2}\in(0.0,\:0.15]\:[\mathrm{kg}]$
(for $m_{i2}\in(0.15,\:1.0]\:[\mathrm{kg}]$, solutions are stable).
As can be easily predicted, the pitchfork bifurcation is no more present
and we observe two disconnected branches of periodic solutions (the
imperfect pitchfork bifurcation). As one can see, 
the maximum amplitudes of the lower pendula start to diverge close
to the destabilization and the second lower pendulum has nearly twice a higher 
amplitude than the first one in the Neimark-Saker bifurcation point 
($m_{12}=m_{22}=0.103\:[\mathrm{kg}]$).
Close to the Neimark-Saker bifurcation located on the main branch,
one can observe an appearance of the second branch which starts and disappears
in saddle-node bifurcations. The stable part of this branch is bounded
by the saddle-node ($m_{12}=m_{22}=0.1019\:[\mathrm{kg}]$) and the
Neimark-Sacker ($m_{12}=m_{22}=0.1047\:[\mathrm{kg}]$) bifurcations
and there is a similar difference in amplitudes between the lower and
upper pendula as for the main branch. The stability range of this separated
branch in the two-parameter space is studied in the next subsection. 

\begin{figure}
\begin{centering}
\includegraphics[scale=0.8]{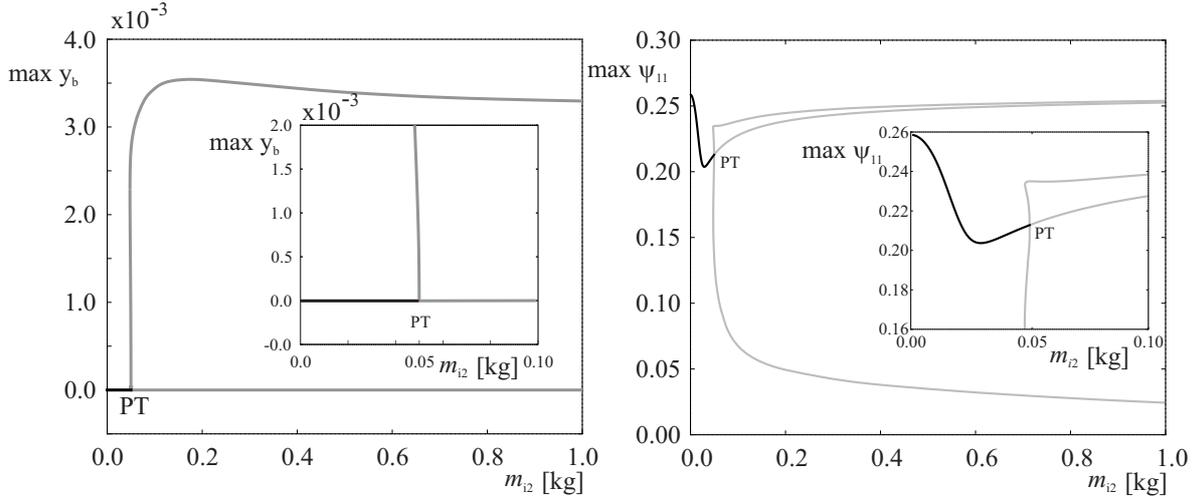}
\par\end{centering}

\caption{One-parameter path-following of the anti-phase synchronous periodic solutions
from Fig. \ref{fig:Idem_PS}(c). A change in the maximum amplitude
of the beam $\mathrm{max}\: y_{b}$ (a) and the second upper pendulum $\mathrm{max}\:\psi_{12}$
(b) is shown for decreasing masses of the lower pendula $m_{i2}=\in\left(0.0,\:1.0\right]\:[\mathrm{kg}]$
($i=1,2$), whereas the upper pendula have masses equal to $1.0\:[\mathrm{kg}]$.
The black and gray lines correspond to stable and unstable periodic
solutions, respectively. The abbreviation PT corresponds to the pitchfork bifurcation.
Bifurcations along unstable branches are neglected.\label{fig:st_sym_c} }
\end{figure}

The same analysis is performed for the periodic solution shown in Fig.
\ref{fig:Idem_PS}(c), which is originally unstable. In Fig. \ref{fig:st_sym_c}(a)
one can see that the maximum amplitude of the beam $\mathrm{max}\; y_{b}$
with decreasing masses of the lower pendula $m_{12}$ and $m_{22}$
remains zero (symmetry is maintained) in the whole range under consideration. For
$m_{12}=m_{22}=0.05\:[\mathrm{kg}]$, we observe the subcritical pitchfork
bifurcation, where symmetric solutions stabilize and stay stable nearly
to $m_{12}=m_{22}\approx0.0\:[\mathrm{kg}]$. The second branch corresponds
to the asymmetric unstable periodic motion with low oscillations of
the beam. All branches coming from the pitchfork bifurcation can be seen
in Fig. \ref{fig:st_sym_c}(b), where we show the maximum amplitude
of the second upper pendulum $\mathrm{max}\;\psi_{12}$. To have a general
overview, we increase the masses $m_{12}$ and $m_{22}$ but the stability
properties do not change, hence the periodic solutions along all branches
stay unstable.

\subsection{Ranges of stability of synchronous solutions in the two-parameter space}

In this subsection, we show how asymmetric changes of pendulum
masses influence the stability of the previously present periodic solutions.
In all cases we start with the pendulum configuration obtained for
the identical double pendula (configurations from Fig. \ref{fig:Idem_PS}(a-d)).
We change the mass of the second upper pendulum $m_{21}$ (in different
intervals for each periodic solutions) and the masses of the lower
pendula $m_{12}=m_{22}$ in the interval $\left(0.0,\:1\right]\,[\mathrm{kg}]$.
Our calculations are presented on the two-dimensional bifurcation
diagrams (masses of the lower pendula $m_{12}=m_{22}$ versus the
mass of the second upper pendulum $m_{21}$). 

\begin{figure}
\begin{centering}
\includegraphics[scale=0.8]{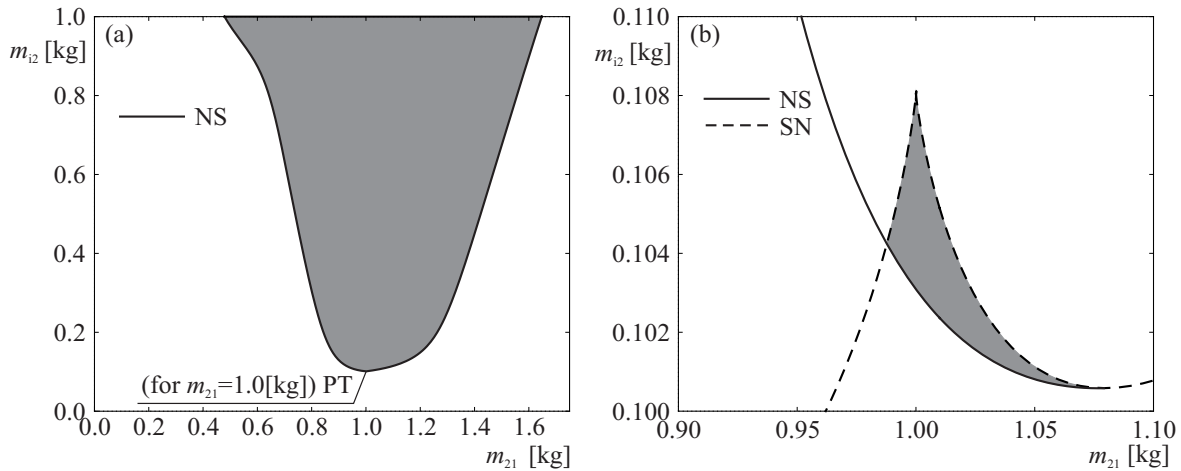}
\par\end{centering}

\caption{In (a) two-parameter continuation of the anti-phase synchronization periodic
solution (the beam is at rest - see Fig. \ref{fig:Idem_PS}(d)) for the masses
$m_{21}$ and $m_{i2}\in(0.0,\:1.0]\:[\mathrm{kg}]$. The observed stable
periodic solutions destabilize thought the Neimark-Sacker bifurcations.
When the system is symmetric ($m_{21}=1.0\:[\mathrm{kg}]$), the Neimark-Sacker
bifurcation is interchanged by the pitchfork bifurcation. In (b) two-parameter plot for the masses $m_{21}$ and $m_{i2}\in(0.0,\:1.0]\:[\mathrm{kg}]$
of the disconnected branch (see Fig. \ref{fig:PP_anti_phase}(c,d)). Stable
periodic solutions destabilize through the Neimark-Sacker (continuous
line) and saddle-node (dashed line) bifurcations. The gray shaded area
corresponds to the existence of stable periodic solutions, whereas the white
one to the unstable solution. \label{fig:LPandNS_2Par}}
\end{figure}

In Fig. \ref{fig:LPandNS_2Par}(a) we show stability ranges of
the configuration presented in Fig. \ref{fig:Idem_PS}(d). The solution
is bounded by the Neimark-Sacker bifurcation, hence 
we observe an appearance of the quasi-periodic motion outside this range. The bifurcation
scenario which occurs for $m_{21}=1.0\:[\mathrm{kg}]$ is different
from the other ones because of the presence of symmetry. The stability at the bottom
is lost not via the Neimark-Sacker but through the pitchfork bifurcation.
When the symmetry is broken ($m_{21}\neq1.0\:[\mathrm{kg}]$), the
pitchfork bifurcation in no more present but there exists a disconnected
stable range of periodic solutions (coming from the  'second' branch -
see Fig. \ref{fig:PP_anti_phase}(c,d)). The area of existence of
these asymmetric period solutions is presented in Fig. \ref{fig:LPandNS_2Par}(b).
As shown in Fig. \ref{fig:PP_anti_phase}(c,d), the stable range
is bounded by the Neimark-Sacker bifurcation (from the bottom) and the
saddle-node bifurcation line from the top. This area is small and when
the difference of masses of the upper pendula ($m_{11}$ and $m_{21}$)
becomes larger than a few percent, it disappears. 

\begin{figure}
\begin{centering}
\includegraphics[scale=0.8]{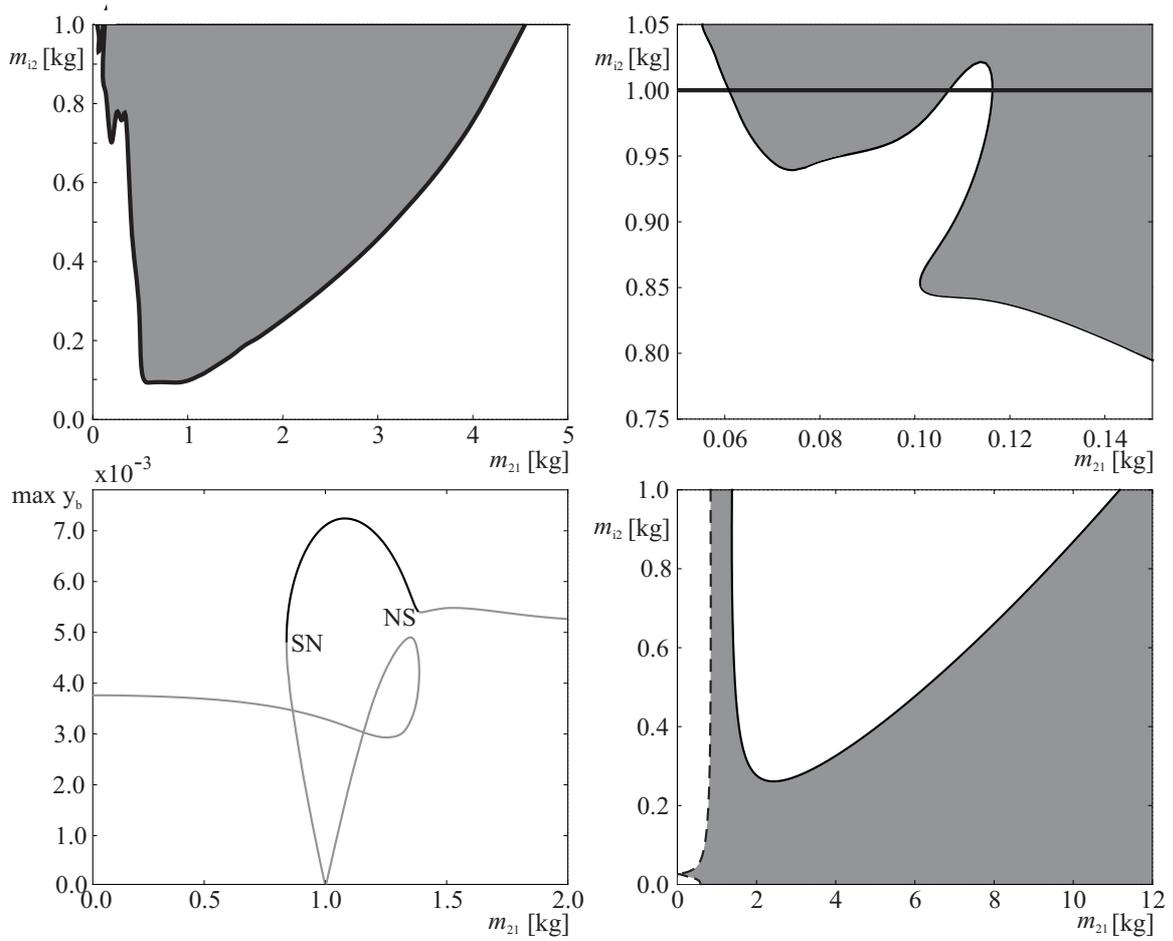}
\par\end{centering}

\caption{In (a,b) two-parameter continuation of the in-phase synchronization periodic
solution (the beam is in the anti-phase state to all pendula) for the masses
$m_{21}$ and $m_{i2}\in(0.0,\:1.0]\:[\mathrm{kg}]$. The observed stable
periodic solutions destabilize through the Neimark-Sacker bifurcation.
In (c) one-parameter ($m_{21}$) plot which shows the connection between
the unstable anti-phase solution (Fig. \ref{fig:Idem_PS}(c)) and
the stable in-phase solution (Fig. \ref{fig:Idem_PS}(b)), where the gray
and black lines correspond to stable and unstable periodic solutions.
Then, in (d) two-parameter plot for the masses $m_{21}$ and $m_{i2}\in(0.0,\:1.0]\:[\mathrm{kg}]$.
Stable periodic solutions destabilize through the Neimark-Sacker (continuous
line) and saddle-node (dashed line) bifurcations. The gray shaded area
corresponds to the existence of stable periodic solutions, whereas the white
one to the unstable solution.\label{fig:all_in_phase}}
\end{figure}

One can distinguish two types of the in-phase motion: the first one where all
pendula are in-phase (Fig. \ref{fig:Idem_PS}(a)) and the second one where
the upper and lower pendula are in-phase but in anti-phase to each other
(Fig. \ref{fig:Idem_PS}(b)). To investigate the first type of motion,
we follow the periodic solution in the two-parameter space (similarly
as for the anti-phase motion). The results of calculations are presented in
Fig. \ref{fig:all_in_phase}(a,b). As can be easily seen, the stable area
is much larger than in the previous case. Solutions destabilize similarly
in the Neimark-Sacker bifurcations, which results in an appearance of
the quasi-periodic motion. For a decreasing mass $m_{21}$, we observe a rapid
jump around $m_{21}=0.5\:[\mathrm{kg}]$ from $m_{i2}\approx0.08\:[\mathrm{kg}]$
to $m_{i2}\approx0.7\:[\mathrm{kg}]$, for an increase in $m_{21}$,
the bound of bifurcation grows nearly linearly reaching $m_{i2}=1.0\:[\mathrm{kg}]$
for $m_{21}=4.6\:[\mathrm{kg}]$. The zoom of the majority of the left part is presented
in \ref{fig:all_in_phase}(b), where one can see that the Neimark-Sacker
bifurcation line has a complex structure. The gap for $m_{i2}$ corresponds
to an appearance of the quasi-periodic motion in the Neimark-Sacker bifurcation
and its disappearance in the inverse Neimark-Sacker bifurcation. 

In Fig. \ref{fig:all_in_phase}(c) we present a one-parameter plot which
shows a connection between the unstable anti-phase solution (Fig.
\ref{fig:Idem_PS}(c)) and the stable in-phase solution (Fig. \ref{fig:Idem_PS}(b)).
The starting solution is the unstable one ($m_{21}=1.0\:[\mathrm{kg}]$
and $\mathrm{max}\: y_{b}=0.0$) with an increasing mass, we do not
observe changes in stability - the unstable branch turns around and
reaches $m_{12}\approx0.0\:[\mathrm{kg}]$. Following the second
direction results in a change of stability in the saddle-node bifurcation
($m_{21}=0.837\:[\mathrm{kg}]$) and then destabilization in the Neimark-Sacker
bifurcation ($m_{21}=1.38\:[\mathrm{kg}]$). For $m_{21}=1.0\:[\mathrm{kg}]$,
the stable solution corresponds to the solution presented in Fig. \ref{fig:Idem_PS}(b).
Next, we follow the bifurcation which bounds the stable branch in the
two-parameter space ($m_{21}$ and $m_{i2}$), which is shown in Fig.
\ref{fig:all_in_phase}(d). The stable range stays narrow up to $m_{i2}\approx0.3\:[\mathrm{kg}]$
where the Neimark-Sacker bifurcation line changes the direction and
starts to go up. When the mass of the upper pendulum is large enough ($m_{21}>11.0\:[\mathrm{kg}]$),
we once again can observe a stable solution for $m_{i2}=1.0\:[\mathrm{kg}]$.
From the left-hand side, as has been mentioned before, the stable area is bounded by
the saddle-node bifurcation line and is nearly a constant line around $m_{21}=0.84\:[\mathrm{kg}].$

\section{Conclusions\label{sec:Conclusions}}

Our studies show that two self-excited double pendula with the van der
Pol type of damping, hanging from the horizontally movable beam, can
synchronize. For identical pendula, four different synchronous configurations
are possible (in-phase or anti-phase), but not all of them are stable
for the given parameters of the beam. When the pendula are nonidentical,
i.e., have different masses, we
observe synchronous states for which the phase difference between
the pendula is close to 0 or $\pi$ for a small parameter mismatch. With an increase in this difference,
we observe a  stable solution with phase shifts between 0 and $\pi$.
They finally destabilize in the Neimark-Sacker saddle-node bifurcations,
which results in an appearance of unsynchronized quasi-periodic oscillations or a jump to another attractor.
Similar synchronous states have been observed experimentally in \cite{ref20}
but a special controlling procedure has been applied to stabilize them.

The observed behavior of system (1) can be explained by the energy
expressions derived in Section 3, which also show why other synchronous
states are not possible. We prove that the observed behavior
of the system is robust as it occurs in a wide range of system parameters.

\section*{Acknowledgment}

This work has been supported by the Foundation for Polish Science,
Team Programme under the project TEAM/2010/5/5.


\section*{References}
\begin{thebibliography}{10}
\bibitem{ref1} A. Andronov, A. Witt, S. Khaikin, \textit{Theory of
Oscillations}. (Pergamon, Oxford 1966)

\bibitem{ref3} I.I. Blekhman, \textit{Synchronization in Science
and Technology.} (ASME, New York 1988).

\bibitem{ref15} A. Pikovsky, M. Roesenblum, J. Kurths,\textit{ Synchronization:
An Universal Concept in Nonlinear Sciences}. (Cambridge University
Press, Cambridge 2001).

\bibitem{ref21} M. P. Aghababa, H. P. Aghababa, Nonlinear Dynamics
\textbf{67}, 2689, (2012).

\bibitem{ref24} M. Kapitaniak, P. Brzeski, K. Czolczynski, P. Perlikowski,
A. Stefanski and T. Kapitaniak, Progress of Theoretical Physics \textbf{128}(6),
1141,(2012).

\bibitem{ref9} C. Huygens, 1893, Letter to de Sluse, In: Oeuveres
Completes de Christian Huygens, (letters; no. 1333 of 24 February
1665, no. 1335 of 26 February 1665, no. 1345 of 6 March 1665), (Societe
Hollandaise Des Sciences, Martinus Nijhoff, La Haye).

\bibitem{ref10} M. Kapitaniak, K. Czolczynski, P. Perlikowski, A.
Stefanski, T. Kapitaniak, Physics Report \textbf{517}, 1, (2012).

\bibitem{ref2} M. Bennet, M.F. Schatz, H., Rockwood, K. Wiesenfeld,
Proc. Roy. Soc. London A \textbf{458}, 563, (2002).

\bibitem{ref4} K. Czolczynski, P. Perlikowski, A. Stefanski, T. Kapitaniak,
Prog. Theor. Phys. \textbf{122}, 1027, (2009).

\bibitem{ref5} K. Czolczynski, P. Perlikowski, A. Stefanski, T. Kapitaniak,
Physica A\textbf{ 388}, 5013, (2009).

\bibitem{ref6} K. Czolczynski, P. Perlikowski, A. Stefanski, T. Kapitaniak,
Chaos\textbf{ 21}, 023129, (2011).

\bibitem{ref7} R. Dilao, Chaos \textbf{19}, 023118 (2009).

\bibitem{ref8} A.L. Fradkov, B. Andrievsky, Int. J. Non-linear Mech.\textbf{
42}, 895, (2007).

\bibitem{ref11} A. Yu. Kanunnikov , R.E. Lamper, J. Appl. Mech \&
Theor. Phys. \textbf{44}, 748, (2003).

\bibitem{ref13} J. Pantaleone, Am. J. Phys. \textbf{70}, 992 (2002).

\bibitem{ref14} P. Perlikowski, M. Kapitaniak, K. Czolczynski, A.
Stefanski, T. Kapitaniak, Int. J. Bif. Chaos \textbf{22}, 1250288
(2012)

\bibitem{ref17} M. Senator, Journal Sound and Vibration, \textbf{291},
566, (2006).

\bibitem{ref19} H. Ulrichs, A. Mann, U. Parlitz, Chaos \textbf{19},
043120 (2009).

\bibitem{ref35} N. Rott, Z. angew. Math. Phys. \textbf{2l}, 570,
(1970).

\bibitem{ref29} J. Miles, Journal of Applied Mathematics and Physics
(ZAMP) \textbf{36}, (1985)

\bibitem{ref28} A.C. Skeldon, Physics Letters A \textbf{166}, 224,
(1992)

\bibitem{ref27} A.C. Skeldon, Physica D \textbf{75}, 541, (1994).

\bibitem{ref26} S. Samaranayake, A.K. Bajaj, Nonlinear Dynamics \textbf{4},
605, (1993).

\bibitem{ref31} T. Morbiato, R. Vitaliani, A. Saetta, Computers \&
Structures \textbf{89}, 1649, (2011).

\bibitem{ref34} A.P. Willmott, J. Dapena, Journal of Sports Sciences
\textbf{30}, 369, (2012).

\bibitem{ref33} Y. Suzukia, T. Nomuraa, M. Casadiob, P. Morassoc,
Journal of Theoretical Biology \textbf{310}, 55, (2012).

\bibitem{ref32} K.P. Granataa, S.E. Wilsonb, Clinical Biomechanics,
Volume \textbf{16}, Issue 8, Pages 650--659 (2001).

\bibitem{ref20} A. Fradkov, B. Andrievsky, K. Boykov, Mechatronics
\textbf{15}, 1289, (2005).

\bibitem{refAuto} E.J. Doedel, A.R. Champneys, T.F. Fairgrieve, Y.A.
Kuznetsov, B. Sandstede, X. Wang\textit{ Auto 97: continuation and
bifurcation software for ordinary differential equations} ( 1998)\end{thebibliography}
\end{document}